\documentclass{article}
\usepackage{amsfonts,amsmath,amsthm,amssymb,authblk,latexsym,bm}
\usepackage{hyperref} \hypersetup{colorlinks=true} 
\usepackage{citeref}

\newtheorem{theorem}{Theorem}[section]
\newtheorem{corollary}[theorem]{Corollary}

\newtheorem{lemma}[theorem]{Lemma}
\newtheorem{definition}[theorem]{Definition}

\newtheorem{remark}[theorem]{Remark}

\numberwithin{equation}{section}

\begin{document}
\title{Dihedral Group Codes over Finite Fields}
\author{
Yun Fan\par
{\small School of Mathematics and Statistics}\par\vskip-1mm
{\small Central China Normal University, Wuhan 430079, China}
\par\vskip2mm
Liren Lin\par
{\small School of Optical Information and Energy Engineering}\par\vskip-1mm
{\small School of Mathematics and Physics}\par\vskip-1mm
{\small Wuhan Institute of Technology, Wuhan 430205, China}}
\date{}
\maketitle

\insert\footins{\footnotesize{\it Email address}:
yfan@mail.ccnu.edu.cn (Yun Fan);
l\_r\_lin86@163.com (Liren Lin).}

\begin{abstract}
Bazzi and Mitter \cite{BM} showed that 
binary dihedral group codes are asymptotically good.
In this paper we prove that the dihedral group codes over any finite field
with strong duality property are asymptotically good. 
If the characteristic of the field is even,
self-dual dihedral group codes are asymptotically good.
If the characteristic of the field is odd, 
maximal self-orthogonal dihedral group codes
and LCD dihedral group codes are asymptotically good. 

\end{abstract}

\section{Introduction}
Let $F$ be a finite field with cardinality $|F|=q$, where $q$ is
 a power of a prime (just the characteristic ${\rm char}\,F$ of $F$).
Let $n$ be a positive integer. Any nonempty subset 
$C\subseteq F^n$ is called a code of length $n$ over $F$ in coding theory. 
The {\em Hamming weight} ${\rm w}(a)$ for $a=(a_1,\cdots,a_n)\in F^n$
is defined to be the number of the indexes $i$ that $a_i\ne 0$, and
the {\em Hamming distance} ${\rm d}(a,b)={\rm w}(a-b)$ for $a,b\in F$.
And ${\rm d}(C)=\min\{{\rm d}(c,c')\,|\,c\ne c'\in C\}$ 
is said to be the {\em minimum distance} of $C$,
while $\Delta(C)=\frac{{\rm d}(C)}{n}$ is called the
{\em relative minimum distance} of~$C$.
The rate of the code $C$ is defined as ${\rm R}(C)=\frac{\log_q|C|}{n}$.
If $C$ is a {\em linear code}, i.e., a linear subspace of $F^n$, then
${\rm R}(C)=\frac{\dim_F C}{n}$.
A class of codes is said to be {\em asymptotically good} if
there is a code sequence $C_1,C_2,\cdots$ in the class such that
the length $n_i$ of $C_i$ goes to infinity and both
the rate ${\rm R}(C_i)$ and the relative minimum distance
$\Delta(C_i)$ are positively bounded from below.

Gilbert \cite{G} and Varshamov \cite{V} showed that,
for linear codes whose relative minimum distances are at least $\delta$, 
 $0<\delta<1-q^{-1}$, 
their rates attain the {\em GV-bound} $g_q(\delta)=1-h_q(\delta)$ with high probability,
where
\begin{equation}\label{h_q}
 h_q(\delta)=\delta\log_q(q-1)-\delta\log_q\delta-(1-\delta)\log_q(1-\delta), 
 \quad  0\le\delta\le 1-q^{-1},
\end{equation}
is the  $q$-{\em entropy} function.
Note that 
$h_q(\delta)$ is increasing and concave in the interval $[0,1-q^{-1}]$. 
More precisely, for linear codes of rate $r$,  
Pierce~\cite{P67} proved that their relative minimum distances 
 are asymptotically distributed at $g_q^{-1}(r)$,
where $g_q^{-1}(\cdot)$ is the inverse function of the GV-bound $g_q(\cdot)$. 
In particular, linear codes are asymptotically good.
For codes of rate $r$, Barg and Forney \cite{BF} showed that 
their relative minimum distances are asymptotically distributed at $g_q^{-1}(2r)$.

Mathematical structures afforded by codes are useful for theory and practice.
The euclidean inner product of $F^n$ is defined as:  
\begin{equation}\label{inner product}\textstyle
\big\langle a,\,b\big\rangle =\sum_{i=1}^n a_ib_i,\quad
\forall~ a=(a_1,\cdots,a_n),\, b=(b_1,\cdots,b_n)\in F^n.
\end{equation} 
And $C^\bot=\{ a\in F^n\;|\;\langle c,a\rangle =0,~\forall~ c\in C\}$
is the {\em orthogonal code} of $C$. 
If $C\subseteq C^\bot$  ($C= C^\bot$, resp.), then
$C$ is said to be {\em self-orthogonal} ({\em self-dual}, resp.).
Obviously, ${\rm R}(C)=\frac{1}{2}$ if $C$ is self-dual.
If $C$ is self-orthogonal, 
but any code containing $C$ properly is not self-orthogonal,
then $C$ is said to be {\em maximal} self-orthogonal.
On the other hand, $C$ is said to be a
{\em linear complementary dual code},
 or {\em LCD code} in short, if $C\bigcap C^\bot=0$.
 
Let $G$ be a finite group, and $FG$ be the 
{\em group algebra} of $G$ over the field~$F$.
Any left ideal of $FG$ is called a {\em group code} of $G$ over $F$, 
or an {\em $FG$-code} for short. 
Further, any $FG$-submodule of $(FG)^2=FG\oplus FG$ is called 
a {\em quasi-$FG$ code of index $2$}, or {\em $2$-quasi} $FG$-code in short. 
Quasi-$FG$ codes of index $m$
are defined similarly. If $G$ is abelian (cyclic, resp.),
quasi-$FG$ codes are also called 
{\em quasi-abelian} codes ({\em quasi-cyclic} codes, resp.) 

Let $G$ be a cyclic group of order $n$. Then $FG$-codes are well-known as 
{\em cyclic codes} of length n over F, 
which are studied and applied extensively since the l950’s.
Even so, it is still an open problem: whether or not the cyclic codes are
asymptotically good? e.g., see \cite{MW06}.
In contrast,
 the quasi-cyclic codes of index $2$ were proved asymptotically good,
 see \cite{CPW, C, K}. 
Moreover, self-dual quasi-cyclic codes are asymptotically good, see \cite{D,LS}.

Now assume that $G$ is a {\em dihedral group} of order $2n$, i.e., $G$   
has a normal cyclic subgroup
$H=\langle u\rangle$ of order $n$ generated by $u$, 
and an element $v$ of order $2$ such that $vuv^{-1}=u^{-1}$.
Then $FG$-codes are called {\em dihedral group codes},
or {\em dihedral codes} in short. 
Bazzi and Mitter \cite{BM} proved that, if $q=2$, 
the binary dihedral codes are asymptotically good.
Their arguments are based on a result in \cite{M74,P85,S86}, 
which estimates the number of the code words in a binary {\em balanced code}
with weight bounded above, see Definition \ref{def balanced} and
Lemma \ref{at most delta} below for details.
Soon after, Mart\'inez-P\'erez and Willems \cite{MW} showed that
the binary doubly-even (hence must be self-dual) 
quasi-cyclic codes of index $2$ are asymptotically good.

In \cite{FL} we generalized the result in \cite{M74,P85,S86} 
on estimating the number of the code words with bounded weight 
 to any q-ary balanced codes, 
see Lemma~\ref{at most delta} below for details; 
and showed that, like the linear codes, 
the relative minimum distances of the quasi-abelian codes 
of rate $r$ are asymptotically distributed at $g^{-1}_q(r)$,
where $g_q^{-1}(\cdot)$ is the inverse function of the GV-bound $g_q(\cdot)$,
see the outline around Eq.\eqref{h_q}.
In that paper we also said  
``... from it (means the generalization Lemma~\ref{at most delta}) 
quite a part of \cite{BM} can be extended to any q-ary case''.

For the case that~${\rm char}\,F\!=\!2$,
Alahmadi, \"Ozdemir and Sol\'e \cite{AOS} discovered~an interesting fact: 
 the self-dual double circulant codes over $F$, a family of
self-dual quasi-cyclic codes of index $2$,  
are in particular self-dual dihedral codes. 
Based on {\em Artin's primitive root conjecture}, 
they proved that such codes are asymptotically good.

For odd $p={\rm char}\,F$, Borello and Willems \cite{BW} considered 
the {\em semidirect products} of the cyclic group of order $p$ 
by suitable finite cyclic groups; with the help of 
the generalization Lemma~\ref{at most delta},
they proved the asymptotic goodness of such group codes.

In this paper we 
extend the asymptotic goodness of dihedral codes to any q-ary case. 
Specifically, we exhibit two kinds of random dihedral codes 
with strong duality property, and with nice asymptotic behavior as well. 
Suitably choosing a positive real number $\delta$ 
and the code lengths $2n_1, 2n_2, \cdots$ going to infinity, 
we prove that the probability for the relative minimum distance 
of the random dihedral codes greater than $\delta$ is convergent to $1$. 
As consequences, we get the following asymptotic goodness. 

\begin{theorem}\label{thm1}
Assume that $0<\delta<1-q^{-1}$ and $0<h_q(\delta)<\frac{1}{4}$.
If ${\rm char}\,F=2$,  then there are self-dual dihedral group codes
$C_1,C_2,\cdots$ over $F$  with length of $C_i$ going to infinity
such that 
$\Delta(C_i)>\delta$ for all $i=1,2,\cdots$.
\end{theorem} 

 For the case ${\rm char}\,F$ is odd,
 \cite{W} had shown that there exist no self-dual dihedral codes. 

\begin{theorem}\label{thm2}
Assume that $0<\delta<1-q^{-1}$ and $0<h_q(\delta)<\frac{1}{4}$.
If ${\rm char}\,F$ is odd, then:

{\bf(1)} 
there are maximal self-orthogonal dihedral group codes
$C_1,C_2,\cdots$ over~$F$ with length of~$C_i$ going to infinity
such that~$\lim\limits_{i\to\infty}{\rm R}(C_i)=\frac{1}{2}$ 
and~$\Delta(C_i)>\delta$ for all $i=1,2,\cdots$. 

{\bf(2)}  
there are LCD dihedral group codes
$C_1,C_2,\cdots$ over~$F$  with length of~$C_i$ going to infinity
such that~${\rm R}(C_i)=\frac{1}{2}$ 
and~$\Delta(C_i)>\delta$ for all~$i=1,2,\cdots$. 
\end{theorem}

If we ignore the action of the element of order $2$ on the normal cyclic 
subgroup of order $n$, then the dihedral group codes 
are quasi-cyclic codes of index $2$ (the converse is not true in general).
So we have consequences:

\begin{corollary} 
{\bf(1)} If ${\rm char}\,F\!=\!2$, then the self-dual quasi-cyclic codes 
of index~$2$ over~$F$ are asymptotically good.

{\bf(2)} If ${\rm char}\,F$ is odd, then 
the maximal self-orthogonal quasi-cyclic codes of index $2$ 
and the LCD quasi-cyclic codes of index $2$ are both asymptotically good.
\end{corollary}

In the next section we sketch preliminaries. 
In \S\ref{D-algebras} we explore the properties 
of the dihedral group algebras over $F$. 
In \S\ref{D-codes} we construct precisely our dihedral group codes 
of length $2n$ with rate $\frac{1}{2}\!-\!\frac{1}{2n}$ or $\frac{1}{2}$;
the two kinds of dihedral group codes may have different behavior. 
In \S\ref{random D-codes} and \S\ref{random D-codes 1/2} 
we exhibit the random properties of the two kinds of dihedral group codes 
constructed in \S\ref{D-codes}. 
The two theorems listed above will be proved 
in \S\ref{proof of thms}.

\section{Preliminaries}\label{preliminaries}

In this paper $F$ is always a finite field with $|F|=q$
which is a power of a prime,
where $|S|$ denotes the cardinality of any set $S$. 
And $n>1$ is an integer.

For any index set $I=\{i_1,\cdots,i_d\}$, 
$F^I=\{(a_{i_1},\cdots,a_{i_d})\,|\, a_{i_j}\in F\}$
is a vector space over $F$ of dimension $d$.
As usual, $F^n=F^I$ with $I=\{1,2,\cdots,n\}$.
For $a\in F^n$, the fraction $\frac{{\rm w}(a)}{n}$ 
is called the {\em relative weight} of $a$.
Let $\delta$ be a real number such that $0<\delta<1-q^{-1}$.
For any code $C\subseteq F^n$, we denote 
$$\textstyle C^{\le\delta}
 =\big\{c\;\big|\;c\in C,~\frac{{\rm w}(c)}{n}\le\delta\big\}.
$$

\begin{definition}\label{def balanced}\rm
Let $C\subseteq F^n=F^I$ where $I=\{1,2,\cdots,n\}$.
If there are subsets $I_1,\cdots,I_s$ (with repetition allowed)
of the index set $I$ and integers $k$ and $t$ such that 
every cardinality $|I_j|=k$ and the following two hold:

{\bf(1)} 
for any $i\in I$, the number of such subscripts
$j$ that $i\in I_j$ is equal to $t$;

{\bf(2)}  
for any $j\!=\!1,\cdots,s$,
the projection $\rho_j\!:F^I\to F^{I_j}$ 
maps $C$ bijectively onto~$F^{I_j}$;

\noindent
then, following \cite{BM} and \cite{P85}, 
we say that $C$ is a {\em balanced code} over $F$ of length $n$ and
{\em information length}~$k$, and $I_1,\cdots,I_s$ form
a {\em balanced system of information index sets} of $C$.
\end{definition}

Note that the phrase ``balanced codes'' might be used for
different concepts in literature, e.g., in \cite{IW}. 
And, in notation of the theory of block designs, 
the above definition is equivalent to saying that 
``there is a $1$-$(n,k,t)$ design whose blocks are information sets''
(thanks are given to the reviewers for showing the concise version).

The following result was proved in \cite{M74}, \cite{P85} and \cite{S86} 
for binary case, and in~\cite{FL} for the present version.

\begin{lemma}\label{at most delta}
Let $C$ be a balanced code over $F$ of length $n$ and information length $k$.
Assume that $0<\delta<1-q^{-1}$.
Then $|C^{\le\delta}|\le q^{kh_q(\delta)}$.
\end{lemma}

If $C$ is a  linear code, 
then ${\rm w}(C)=\min\{{\rm w}(c)\;|\;0\ne c\in C\}$ 
is called the {\em minimum weight} of~$C$, and ${\rm w}(C)={\rm d}(C)$. 
So $\Delta(C)=\frac{{\rm w}(C)}{n}$, and it is also 
called the {\em relative minimum weight} of~$C$.
And the rate $R(C)=\frac{\dim_F C}{n}$.

Let $G$ be a finite group, 
$FG=\big\{\sum_{x\in G}a_x x\:\big|\; a_x\in F\big\}$,
which is an $F$-vector space with a multiplication induced 
by the multiplication of the group~$G$. 
So $FG$ is an $F$-algebra, called the {\em group algebra} of $G$ over $F$.
Any $\sum_{x\in G}a_x x\in FG$ is viewed as a sequence $(a_x)_{x\in G}$
of $F$ indexed by $G$.
Any left ideal $C$ of $FG$ is called a {\em group code} of $G$ over $F$.
We also say that $C$ is an $FG$-code for short.
If $e\in FG$ is an {\em idempotent}, i.e., $e^2=e$, 
then $FGe$ is a left ideal and $FG=FGe\oplus FGe'$, 
where $e'=1-e$ is also an idempotent and $ee'=e'e=0$.   
Further, if the idempotent $e$ is central, then $FG=FGe\oplus FGe'$
with both $FGe$ and $FGe'$ being $2$-sided ideals. 
If the greatest common divisor $\gcd(|G|,q)=1$, 
then any ideals and any left ideals can be 
constructed by idempotents in this way;
and $e$ is called a {\em primitive} idempotent once 
$FG e$ is a minimal left ideal (i.e., any left ideal
contained in $FGe$ is either $0$ or $FGe$ itself). 
Please see \cite[Chapter 5, \S3]{HB}, 
or see \cite[\S 4.3]{HP} for cyclic codes.

\begin{remark}\label{g-code balanced}\rm
Any group code $C$ of the group algebra $FG$ is a balanced code, 
see \cite[Lemma 2.2]{BM}. In fact, it can be proved in a similar way
that any transitive permutation codes are balanced codes
(a linear code is called a transitive permutation code if there is a
group permuting the bits of the code transitively and the code
is invariant under the group action, see \cite{FY}). 
\end{remark}

Mapping $x$ to $ x^{-1}$ 
is an anti-automorphism of the group $G$, 
where~$x^{-1}$ denotes the inverse of $x$.  We have
an anti-automorphism of the algebra $FG$: 
\begin{equation}\label{bar map}\textstyle
 FG\longrightarrow FG,~~ \sum\limits_{x\in G}a_x x
 \longmapsto \sum\limits_{x\in G}a_x x^{-1}.
\end{equation}
We denote $\sum\limits_{x\in G}a_x x^{-1}
 =\overline{\sum\limits_{x\in G}a_x x}$,   
and call Eq.\eqref{bar map} the ``bar'' map of $FG$ for convenience. 
So, $\overline{\overline a}=a$, $\overline{ab}=\overline b\,\overline a$,
for $a,b\in FG$. It is an automorphism of $FG$ 
once $G$ is abelian. The following is a linear form of $FG$:
$$\textstyle
\sigma:~ FG\longrightarrow F,~~ 
\sum\limits_{x\in G}a_x x\longmapsto a_{1_G} ~~~
(\mbox{$1_G$ is the identity of $G$}).
$$

For $a,b\in FG$,
we use the notation $\langle a,b\rangle$ to denote the euclidean inner product of
 $a$ and $b$, which are viewed as sequences   
$(a_x)_{x\in G}$ and $(b_x)_{x\in G}$ of length $n$ over $F$;
see Eq.\eqref{inner product}.

\begin{lemma}\label{sigma bar}
{\bf(1)} $\sigma(ab)=\sigma(ba)$, $\forall~a,b\in FG$.

{\bf(2)} $\langle a, b\rangle=\sigma(a\overline b)=\sigma(\overline a b)$, 
          $\forall~a,b\in FG$.

{\bf(3)} $\langle d\,a,\,b\rangle=\langle a,\,\overline d\, b\rangle$, 
          $\forall~a, b,d\in FG$.

{\bf(4)} If $C$ is an $FG$-code, then so is $C^\bot$.

{\bf(5)} For $FG$-codes $C$ and $D$, 
 $\langle C,D\rangle=0$ if and only if $C\overline D=0$.
\end{lemma}

{\it Proof}.~
The (1), (2) is verified directly. The (3) follows from (2). And (4) is checked by (3).
For (5), the sufficiency follows from (2) directly. Conversely, 
if $c\,\overline d\ne 0$ for $c\in C$ and $d\in D$,
write $c\,\overline d=\sum\limits_{x\in G}b_x x$ 
with a coefficient $b_{x_0}\ne 0$;
then $x_0^{-1}c\in C$ and $\big\langle x_0^{-1}c,d\big\rangle
=\sigma\big(x_0^{-1}c\,\overline d\big)=b_{x_0}\ne 0$. 
\qed 

\medskip
Assume that $H$ is a cyclic group of order $n$. 
Then $FH$-codes are cyclic codes, 
and can be described by monic factors of the polynomial $X^n-1$.
In the following, we further assume that $\gcd(n,q)=1$.
Then monic factors of $X^n-1$ are determined by their zeros. 
As noted above,  $FH$-codes are determined by idempotents.
So each ideal $FHe$ with $e^2=e\ne 0$ corresponds to a monic factor
$g(X)\,\big|\,X^n-1$ such that $FHe\cong F[X]/\langle g(X)\rangle$. 
If the ideal $FHe$ is simple, i.e., $e$ is a primitive idempotent,
then $g(X)$ is irreducible and $FHe$ is a field over~$F$ with
extension degree $\dim_F FHe=\deg g(X)$.
Thus $FH$ has finitely many primitive idempotents
$e_0,e_1,\cdots,e_s$ such that $1=e_0+e_1+\cdots+e_s$
and $e_ie_j=0$ for $0\le i\ne j\le s$,
where $e_0=\frac{1}{n}\sum_{x\in H}x$ and $\dim_F FHe_0=1$.
And the automorphism ``bar'' in Eq.\eqref{bar map} 
permutes the primitive idempotents.

For any ring (with identity) $R$, by $R^\times$ we denote the multiplicative group
consisting of the units (invertible elements) of $R$.
By ${\Bbb Z}_n$ we denote the integer residue ring modulo $n$, 
hence ${\Bbb Z}_n^\times$ is the multiplicative group consisting of 
the reduced residue classes.  Then $q\in{\Bbb Z}_n^\times$ (since $\gcd(n,q)=1$).
In the multiplicative group ${\Bbb Z}_n^\times$,
${\rm ord}_{{\Bbb Z}_n^\times}(q)$ denotes the order of $q$,
and $\big\langle q\big\rangle_{{\Bbb Z}_n^\times}$
denotes the cyclic subgroup generated by $q$.
The following facts are well-known. 

\begin{lemma} \label{n coprime to q}
Let $H$ be a cyclic group of odd order $n$ with  
$\gcd(n,q)\!=\!1$. 
Let $e_0,e_1,\cdots,e_s$ be all primitive idempotents of $FH$, 
where $e_0=\frac{1}{n}\sum_{x\in H}x$. Let
$\lambda(n)=\min\big\{\dim_F(FHe_1),\cdots, \dim_F(FHe_s)\big\}$.

{\bf(1)}\,{\rm(\cite[Lemma 2.5]{BM})}\, 
$\lambda(n)=\min\{ {\rm ord}_{{\Bbb Z}_{p}^\times}(q)\;|\;
\mbox{$p$ is a prime divisor of $n$}\}$.
 
{\bf(2)}\,{\rm(\cite[Theorem 6]{AKS})}\, 
$\overline e_j\ne e_j$ for any $j>0$ if and only if
${\rm ord}_{{\Bbb Z}_n^\times}(q)$ is odd.

{\bf(3)}\,{\rm(\cite[Theorem 1]{KR})}\,
$\overline e_j=e_j$ for any $j>0$ if and only if
$-1\in\big\langle q\big\rangle_{{\Bbb Z}_n^\times}$.
\end{lemma}

We need some number-theoretic results. Let $t>q$ be an integer,
and $\pi(t)$ be the number of the primes less or equal to $t$.
By Gauss' Lemma, $\lim\limits_{t\to\infty}\frac{\pi(t)}{t/\ln t}=1$.

\begin{lemma}\label{good primes}
Set ${\cal G}_t\!=\!\big\{\mbox{\rm prime $p$}\,\big|\, 
  q<p\le t,~ {\rm ord}_{{\Bbb Z}_p^\times}(q)\ge(\log_q t)^2\big\}$.
Then the natural density
$\lim\limits_{t\to\infty}\frac{|{\cal G}_t|}{\pi(t)}=1$.
\end{lemma}

{\it Proof}.~ It was proved in \cite[Lemma 2.6]{BM} for the binary case.
For the general case, the proof is similar.
Set $\overline{\cal G}_t\!=\!\big\{\mbox{\rm prime $p$}\,\big|\, 
 q<p\le t,~{\rm ord}_{{\Bbb Z}_p^\times}(q)<(\log_q t)^2\big\}$.
If $r<(\log_q t)^2$ and $p_1,\cdots,p_k\in \overline{\cal G}_t$ satisfy that
${\rm ord}_{{\Bbb Z}_{p_i}^\times}(q)=r$, $i=1,\cdots,k$, 
then $q^r\!-\!1=p_1\cdots p_k s$, hence $k\le \log_q(q^r\!-\!1)<r<(\log_q t)^2$. 
So $|\overline{\cal G}_t|<(\log_q t)^4$, and
$$\textstyle
\lim\limits_{t\to\infty}\frac{|\overline{\cal G}_t|}{\pi(t)}
 <\lim\limits_{t\to\infty}\frac{(\ln t/\ln q)^4}{t/\ln t} =0.
 \eqno\qed
$$

The following result was proved in \cite{Hasse}
(for  Dirichlet density) and in \cite{O} (for natural density).

\begin{lemma}[\cite{Hasse}, \cite{O}]\label{Hasse, Odoni}
Let ${\cal O}_t=\big\{\mbox{\rm prime $p$}\;\big|\; 
  q<p\le t,~ \mbox{\rm ${\rm ord}_{{\Bbb Z}_p^\times}(q)$ is odd}\big\}$.
Then the natural density 
$\lim\limits_{t\to\infty}\frac{|{\cal O}_t|}{\pi(t)}$
is a positive fraction less than $1$ 
(the exact value depends on the exponent of the prime power $q$,
see \cite[Theorem 1]{O}).
\end{lemma}

With the above three lemmas and their notation, we conclude:

\begin{corollary}\label{infinitely n_i}
{\bf(1)} 
There is a sequence $n_1,n_2,\cdots$ of positive odd integers coprime to $q$
such that ${\rm ord}_{{\Bbb Z}_{n_i}^\times}(q)$ are odd for all $i=1,2,\cdots$
and $\lim\limits_{i\to\infty}\frac{\log_q n_i}{\lambda(n_i)}=0$.   

{\bf(2)}
There is a sequence $n_1,n_2,\cdots$ of positive odd integers coprime to $q$
such that $-1\in\big\langle q\big\rangle_{{\Bbb Z}_{n_i}^\times}$ 
for all $i=1,2,\cdots$ and 
$\lim\limits_{i\to\infty}\frac{\log_q n_i}{\lambda(n_i)}=0$.   
\end{corollary}

{\it Proof}.~ 
(1).~ 
The natural density 
$$\textstyle
\lim\limits_{t\to\infty}\frac{|{\cal O}_t\bigcap{\cal G}_t|}{\pi(t)}
=\lim\limits_{t\to\infty}
 \big(\frac{|{\cal O}_t|}{\pi(t)}+\frac{|{\cal G}_t|}{\pi(t)}
-\frac{|{\cal O}_t\bigcup{\cal G}_t|}{\pi(t)}\big)=
\lim\limits_{t\to\infty}\frac{|{\cal O}_t|}{\pi(t)}>0.$$

(2).~
Note that, if $n$ is a prime, then ${\Bbb Z}_n^\times$ is cyclic
and has $-1$ as the unique element of order $2$.
Hence, 
$-1\in\big\langle q\big\rangle_{{\Bbb Z}_n^\times}$ if and only if
 ${\rm ord}_{{\Bbb Z}_n^\times}(q)$ is even.
Let $\overline{\cal O}_t=\big\{\mbox{\rm prime $p$}\;\big|\; 
  q<p\le t,~ \mbox{\rm ${\rm ord}_{{\Bbb Z}_p^\times}(q)$ is even}\big\}$.
By Lemma \ref{Hasse, Odoni}, the natural density 
$\lim\limits_{t\to\infty}\frac{|{\overline{\cal O}}_t|}{\pi(t)}$ 
is positive. So
$$\textstyle
\lim\limits_{t\to\infty}\frac{|\overline{\cal O}_t\bigcap{\cal G}_t|}{\pi(t)}
=\lim\limits_{t\to\infty}
  \big(\frac{|\overline{\cal O}_t|}{\pi(t)}+\frac{|{\cal G}_t|}{\pi(t)}
-\frac{|{\overline{\cal O}}_t\bigcup{\cal G}_t|}{\pi(t)}\big)=
\lim\limits_{t\to\infty}\frac{|{\overline{\cal O}}_t|}{\pi(t)}>0.
\eqno\qed
$$

\begin{lemma}\label{k_t}
Let $q\ge 2$ and $k_1\le k_2\le\cdots\le k_m$ be positive integers.
If $k_1\ge \log_qm$, then 
$(q^ {k_1}-1)(q^{k_2}-1)\cdot\cdots\cdot (q^{k_m}-1)\ge q^{k_1+k_2+\cdots+k_m-2}.$
\end{lemma}
{\it Proof}. We have
$$\textstyle
 \frac{(q^{k_1}-1)(q^{k_2}-1)\cdots (q^{k_m}-1)}
  {q^{k_1} q^{k_2}\cdots q^{k_m}}
 =(1-\frac{1}{q^{k_1}})(1-\frac{1}{q^{k_2}})\cdots(1-\frac{1}{q^{k_m}})
\ge(1-\frac{1}{q^{k_1}})^m.
$$
Note that the sequence $(1-t^{-1})^t$ for $t=2,3,\cdots$ is increasing and 
$(1-\frac{1}{2})^2\ge\frac{1}{q^2}$.
Since $m\le q^{k_1}$, we get that
$(1-\frac{1}{q^{k_1}})^m\ge
  (1-\frac{1}{q^{k_1}})^{q^{k_1}}\ge \frac{1}{q^{2}}$.
\qed

\section{Dihedral group algebras}\label{D-algebras}
\begin{remark}\label{notation G}\rm
In the following we always assume that:
\begin{itemize}
\item\vskip-3pt
$F$ is a finite field of cardinality $q$.
\item
$n>1$ is an odd integer and $\gcd(n,q)=1$.
\item
$G=\langle u,~v\mid u^n=1=v^2,~ vuv^{-1}=u^{-1}\rangle$
 is the dihedral group of order $2n$.

$H=\langle u\rangle\le G$ is the cyclic subgroup generated by $u$
of order $n$;

$vH=\{v,vu,\cdots,vu^{n-1}\}=Hv$ is the coset of $H$ other than $H$;
 
Hence $G=H\cup vH$.
\item
$FG=\big\{\sum_{x\in G}a_x x\;\big|\; a_x\in F\big\}$
is the group algebra of $G$ over $F$. 
\end{itemize}
\end{remark}

\begin{lemma}\label{basic FG}
$FH$ is a commutative ring, $FG=FH\oplus vFH$,  $vFH=FvH=FHv$, 
 and the following hold.

{\bf(1)} 
Let $e_0\!=\!\frac{1}{n}\sum_{x\in H}x$, 
$e_1,\cdots,e_s$ be all primitive idempotents of $FH$.
Then $FH=FHe_0\oplus FHe_1\oplus \cdots\oplus FHe_s$
is a direct sum of simple ideals $FHe_j$'s
which are field extensions over $F$.
In particular, $FHe_0=Fe_0$ is the trivial ideal with 
${\dim_F FHe_0=1}$. 

{\bf(2)}   
$H$ is normal in $G$, and
$v$ induces the automorphism ``bar'' on $FH$, i.e.,
in notation of Eq.\eqref{bar map}, $vav^{-1}=\overline a$, for all $a\in FH$.

{\bf(3)}  
The idempotent $e_0$ is central in $FG$ and 
the ideal $FGe_0$ is of dimension~$2$. 
Set $\widehat e_0=e_0+ve_0$; then $\widehat e_0$ is central in $FG$,
 $\widehat e_0\kern0.5pt\overline{\widehat e_0}
  ={\widehat e_0}^{\,2}=2\widehat e_0$ and 
$FG\widehat e_0
 =\big\{ a\sum_{x\in G}x\;\big|\; a\in F\big\}=F\widehat e_0
$~is an ideal of dimension $1$ contained in $FGe_0$.
\end{lemma}

{\it Proof}.~ (1) is well-known, see \cite[\S 4.3]{HP}.
The others can be checked straightforwardly. \qed

\medskip
By ${\rm M}_2(F)$ we denote the $2\times 2$ matrix algebra over $F$.

\begin{lemma}\label{FHe}
Let $e$ be a primitive idempotents of $FH$ other than $e_0$. Then
$FHe$ is a field extension over $F$, 
and one of the following holds:

{\bf(1)}  
If $\overline e\ne e$, 
then $e+\overline e$ is a primitive central idempotent of $FG$, 
and the ideal 
 $FG(e+\overline e)
  =FHe\oplus FH\overline e\oplus vFHe\oplus vFH\overline e
\cong{\rm M}_2(\widetilde F)$, where $\widetilde F=FHe$.

{\bf(2)}  
If $\overline e=e$, then $e$ is a primitive central idempotent of $FG$,
the extension degree $\dim_F FHe$ is even,
$FHe$ has a subfield $\widetilde F$ with $\dim_{\widetilde F}FHe=2$,
and the ideal  $FGe=FHe\oplus FHev\cong{\rm M}_2(\widetilde F)$.
\end{lemma}

{\it Proof}.~ They are somewhat known, e.g., 
(2) is proved in \cite{BM} for binary case. 
We show a proof for (1), (2) by constructing specific isomorphisms 
\eqref{FHe=M_2}, \eqref{FGe cong}
for later quotation.  

We have seen in Lemma \ref{basic FG}(1) that
$FHe$ is a field.

(1).~ Since $e\overline e=0$, $e+\overline e$ is an idempotent. 
By Lemma \ref{basic FG}(2),   
$v(e+\overline e)=\overline ev+ev=(e+\overline e)v$,
i.e., $e+\overline e$ is central in $FG$.
So 
$$
  FG(e+\overline e)=(FH\oplus vFH)(e+\overline e)
  =FHe\oplus FH\overline e\oplus vFHe\oplus vFH\overline e
$$ 
is an ideal of $FG$. 
Note that $a=ae$ for $a\in\widetilde F=FHe$. 
Define a map:
\begin{equation}\label{FHe=M_2}
\begin{array}{rcl}
{\rm M}_2(\widetilde F)&{\longrightarrow}
 & FHe\oplus vFHe\oplus vFH\overline e\oplus FH\overline e,\\[3pt]
 \begin{pmatrix}a_{11}&a_{12}\\ a_{21}& a_{22} \end{pmatrix}
 &\longmapsto &
a_{11}e+va_{21}e+v\overline{a_{12}}~\overline{e}+\overline{a_{22}}~\overline{e},
\end{array}\end{equation}
which is obviously a linear isomorphism. Note that 
$e\overline e=0$, $ev=v\overline e$ and $ve=\overline e v$, 
see Lemma \ref{basic FG}(2). 
For any two elements of $FG(e+\overline e)$:
$$
a_{11}e+va_{21}e+v\overline{a_{12}}~\overline{e}
 +\overline{a_{22}}~\overline{e},
\quad
b_{11}e+vb_{21}e+v\overline{b_{12}}~\overline{e}
+\overline{b_{22}}~\overline{e},
$$
where $a_{ij},b_{ij}\in \widetilde F$ for $1\le i,j\le 2$,
\begin{eqnarray*}
&&\big(a_{11}e+va_{21}e+v\overline{a_{12}}~\overline{e}
 +\overline{a_{22}}~\overline{e}\big)
\big(b_{11}e+vb_{21}e+v\overline{b_{12}}~\overline{e}
 +\overline{b_{22}}~\overline{e})\\
&=& (a_{11}b_{11}+a_{12}b_{21})e+v(a_{21}b_{11}+a_{22}b_{21})e\\
&&\quad +v\overline{(a_{11}b_{12}+a_{12}b_{22})}~\overline e
 ~+~\overline{(a_{21}b_{12}+a_{22}b_{22})}~\overline e.
\end{eqnarray*}
Thus Eq.\eqref{FHe=M_2} is an algebra isomorphism.

(2).~
In this case, $ve=\overline ev=ev$, hence $e$ is central in $FG$.
Denote $K=FHe$ which is a field with identity $e$.
Note that $n>1$ is odd, see Remark \ref{notation G}.
The map $a\mapsto \overline a$ for $a\in K$  
is an automorphism of the field $K$ of order $2$.
By Galois Theory, 
$$
\widetilde F:=\big\{a\;\big|\; a\in K,~\overline a=a\big\}\subseteq K
  ~~\mbox{is a subfield and}~~  |K:\widetilde F|=2. 
$$
Since $FH=\sum_{i=0}^{n-1}Fu^i$, 
$K=FHe=\sum_{i=0}^{n-1}F(ue)^i$;
i.e., $K$ is generated as an $F$-algebra by~$ue$.  
Then  $K$ is generated as an $\widetilde F$-algebra also by $ue$
(since $\widetilde F\supseteq F$),
and the minimal polynomial $\varphi_{ue}(X)$ 
over $\widetilde F$ of $ue$ has degree $2$.  
Let $\varphi_{ue}(X)=X^2+gX+h\in\widetilde F[X]$.
Then $K=\widetilde F e+\widetilde F(ue)$, and
 $(ue)^2+g(ue)+h=0$. 
Hence
$(\overline{ue})^2+g(\overline{ue})+h=\overline{(ue)^2+g(ue)+h}=0$. 
So $ue$ and $\overline{ue}$ are two roots 
($\overline{ue}\ne ue$ since $ue\notin\widetilde F$) of 
$\varphi_{ue}(X)$, hence $(ue)(\overline{ue})=h$. In $K$ we have
$\overline{ue}=v(ue)v^{-1}=vuv^{-1}e=u^{-1}e=(ue)^{-1}$.
Thus 
$h=(ue)(\overline{ue})=1$, and $ \varphi_{ue}(X)=X^2+gX+1$.
Note that, since $\varphi_{ue}(X)$ is irreducible, 
$g$ and $2$ are not both zero. 
We set 
\begin{equation}\label{eta nu}
\varepsilon=\begin{pmatrix}1&0\\0&1\end{pmatrix},\quad
 \eta=\begin{pmatrix}-g&1\\-1&0 \end{pmatrix},\quad
  \nu=\begin{pmatrix}-1&0\\-g&1 \end{pmatrix}.
\end{equation}
Then the characteristic polynomial of $\eta$ is
$\varphi_\eta(X)=X^2+gX+1=\varphi_{ue}(X)$.
Mapping $e\mapsto\varepsilon$ and $ue\mapsto\eta$,
we get a field isomorphism
\begin{equation}\label{FHe cong}
K=\widetilde F e+\widetilde F(ue)
 \cong \widetilde F[X]/\langle\varphi_{\eta}(X)\rangle
\cong \widetilde F\varepsilon+\widetilde F\eta
\subseteq{\rm M}_2(\widetilde F).
\end{equation}
Comparing the $K$-dimension, we get that
$$
{\rm M}_2(\widetilde F)=
(\widetilde F\varepsilon+\widetilde F\eta)+
(\widetilde F\varepsilon+\widetilde F\eta)\nu.
$$
On the other hand,
$$FGe=FHe+FHev=K+K(ve)=
 \widetilde F e+\widetilde F ue+ \widetilde F ve+\widetilde F uve.$$
Since 
\begin{equation*}
\nu^2=\varepsilon,\qquad
 \nu\eta\nu^{-1}=
       \begin{pmatrix}0&-1\\1&-g\end{pmatrix}=\eta^{-1},
\end{equation*}
by mapping  $ve\mapsto\nu$, $uve\mapsto\eta\nu$,
we extend the isomorphism Eq.\eqref{FHe cong}
to the following isomorphism (where $a,b,c,d\in\widetilde F$):
\begin{equation}\label{FGe cong}
\begin{array}{rcl}
FGe&{\longrightarrow}
 & {\rm M}_2(\widetilde F),\\[3pt]
 ae +b ue+cve+duve
 &\longmapsto &
a\varepsilon +b \eta+c\nu+d\eta\nu,
\end{array}
\end{equation}
which completes the proof.
\qed

\medskip
Combining Lemma \ref{basic FG} and Lemma \ref{FHe}, 
we obtain the following theorem.

\begin{theorem}\label{FG=} 
The dihedral group algebra $FG$ is a direct sum of ideals $A_t$:
$$FG=A_0\oplus A_1\oplus\cdots\oplus A_m,$$
where $A_0=FG e_0$ is described in Lemma \ref{basic FG}(3)
and, for $t=1,\cdots,m$, the ideal  $A_t\cong {\rm M}_2(F_t)$ with 
$F_t$ being a field extension over $F$ and $\dim_F F_t=k_t$,
hence 
\begin{equation}\label{k_t+}\textstyle
  k_1+\cdots+k_m=\frac{n-1}{2}.
\end{equation} 
For the identity $1_{A_t}$ of $A_t$, which is a central idempotent of $FG$,
one of the following two holds:

{\bf(1)} The identity 
$1_{A_t}=e+\overline e$ for a primitive idempotent $e$ of $FH$ 
with $e\ne\overline e$, and $k_t=\dim_F(FHe)$.

{\bf(2)}  The identity 
$1_{A_t}=e$ is a primitive idempotent of $FH$ 
 with $e=\overline e$, and $k_t=\frac{1}{2}\dim_F(FHe)$.
\end{theorem}

\begin{corollary}\label{k_1 lambda}
For $t=1,\cdots,m$,
we have $2k_t\ge\lambda(n)$. 
\end{corollary}

{\it Proof}.~ Recall from Lemma~\ref{n coprime to q} that
$$\lambda(n)=
 \min\big\{\dim_F FHe \;\big|\; \mbox{$e$ is a primitive idempotent of 
  $FH$ other than $e_0$}\big\}.
$$
By Theorem~\ref{FG=}, 
if $\overline e\ne e$ then $k_t=\dim_F FHe$;
otherwise, $k_t=\frac{1}{2}\dim_F FHe$. That is,
$2k_t\ge\lambda(n)$. 
\qed

\medskip
We collect in the following lemma 
the properties of $2\times 2$ matrix algebras
which we need to study the dihedral group codes. 

\begin{lemma}\label{M algebra}
Let $M={\rm M}_2(F)$,  
$\varepsilon=\begin{pmatrix} 1&0\\ 0&1\end{pmatrix}$,
hence ${\rm Z}(M)=F\varepsilon\cong F$ is the center of $M$.
Then $M$ has a subalgebra $E$ which is a field extension over 
$F$ with $\dim_F E=2$, and the following hold:

{\bf(1)} 
If $c\in M$ has ${\rm rank}(c)=1$,
then $Mc=Ec$ is a simple left ideal of~$M$.

{\bf(2)} 
Let $L=E\varepsilon_{11}=M\varepsilon_{11}$,
where $\varepsilon_{11}=\begin{pmatrix} 1&0\\ 0&0\end{pmatrix}$.
Then, for $0\ne c\in L$ and $a,b\in E^\times$, 
$ac=cb$ if and only if $a=b\in (F\varepsilon)^\times$.

{\bf(3)} 
For $\beta\in E^\times$,  
$L\beta$ is a simple left ideal of $M$. 
And, when $\beta$ runs over $E^\times$,   
the $L\beta$ runs over all the simple left ideals of $M$,
each of them appears exactly $q-1$ times. 
\end{lemma}

{\it Proof}. 
The finite field~$F$ has an extension of degree $2$, in other words, 
there is an irreducible polynomial $\varphi(X)$ of degree $2$ over $F$.
Let $\eta\in M$ be a matrix with characteristic polynomial $\varphi(X)$.
Then
$$
E=F\varepsilon+ F\eta= (F\varepsilon)+(F\varepsilon)\eta
\cong F[X]/\langle\varphi(X)\rangle
$$
 is a field extension over $F$ of degree $2$.

(1). 
Obviously, $Ec\subseteq Mc$ and $Mc$ is a  left ideal of $M$ 
with ${\dim_F Mc=2}$. 
Since $E$ is a field and $Ec\ne 0$, hence 
$\dim_E Ec=1$ and ${\dim_F Ec=2}$. So $Ec=Mc$. 
Since any left $M$-submodule contained in $Mc$ is also a left 
$E$-submodule and $\dim_E Mc=1$, $Mc=Ec$ is a simple left ideal.

(2). 
The sufficiency is obvious. We prove the necessity. 
First assume that $c=\varepsilon_{11}$; 
i.e., $a\varepsilon_{11}=\varepsilon_{11}b$. 
Write 
$\eta=\begin{pmatrix}g_{11}&g_{12}\\ g_{21}&g_{22}\end{pmatrix}$,
then $g_{12}\ne 0\ne g_{21}$, 
otherwise the characteristic polynomial of $\eta$ 
is $(X-g_{11})(X-g_{22})$ which is reducible.
Write $a=a_1\varepsilon+a_2\eta$ and 
$b=b_1\varepsilon+b_2\eta$, where $a_i,b_i\in F$.
Then
$$(a_1\varepsilon+a_2\eta)\varepsilon_{11}=
 \varepsilon_{11}(b_1\varepsilon+b_2\eta),
$$
i.e.,
$$
  (a_1-b_1)\varepsilon_{11}+a_2\eta\varepsilon_{11}
  -b_2\varepsilon_{11}\eta=0;
$$
in matrix version,
$$
 \begin{pmatrix} a_1-b_1+a_2g_{11}-b_2g_{11} & -b_2g_{12} 
 \\ a_2g_{21} & 0  \end{pmatrix}=0.
$$
So $a_2g_{21}=-b_2g_{12}=0$.
Since $g_{12}\ne 0\ne g_{21}$, we obtain that
$a_2=b_2=0$ and $a_1=b_1$; i.e., $a=b\in (F\varepsilon)^\times$.

Next, assume that $0\ne c\in L$ and $ac=cb$. 
Since $L=E\varepsilon_{11}$,
there is a $d\in E^\times$ such that $c=d\varepsilon_{11}$.
So $ad\varepsilon_{11}=d\varepsilon_{11}b$. 
Note that $d^{-1}\in E$ commutes with $a$.
Left multiplying by $d^{-1}$,
we get $a\varepsilon_{11}=\varepsilon_{11}b$.
Thus $a=b\in (F\varepsilon)^\times$.

(3). Because  $\beta$ is  invertible,
the map $L\to L\beta$, $c\mapsto c\beta$, is an isomorphism
of left $M$-modules. 
Hence $L\beta$ is a simple left ideal of~$M$.

Next, for $\beta,\beta'\in E^\times$,
 $L\beta=L\beta'$ if and only if $L=L\beta'\beta^{-1}$.
 Denote $b=\beta'\beta^{-1}\in E$.
 Note that $L=E\varepsilon_{11}$ and $Lb=E\varepsilon_{11}b$.
Hence $L=Lb$ if and only if there is an $a\in E$ such that 
$a\varepsilon_{11}=\varepsilon_{11}b$.
By the above (2), $a\varepsilon_{11}=\varepsilon_{11}b$
if and only if $a=b\in (F\varepsilon)^\times$. We get that

$\bullet$~ {\it For $\beta,\beta'\in E^\times$, 
 $L\beta=L\beta'$ if and only if $\beta'\beta^{-1}\in (F\varepsilon)^\times$.
}

\noindent
Thus, when $\beta$ runs over $E^\times$,  
we obtain altogether $\frac{q^2-1}{q-1}=q+1$
distinct simple left ideals $L\beta$ of $M$,
each of them appears $q-1$ times.
On the other hand,
any simple left ideal of $M$ consists of the zero matrix 
and $q^2-1$ matrices of rank~$1$.
Furthermore,
the intersection  of  any two distinct simple left ideals of $M$ is $0$.
The number of the matrices of rank $2$
is equal to $(q^2-1)(q^2-q)=q^4-q^3-q^2+q$.
Hence the number of the matrices of rank~$1$ is equal to
$$q^4-1-(q^4-q^3-q^2+q)=q^3+q^2-q-1=(q+1)(q^2-1).$$
So the number of the simple left ideals of $M$ is: 
$(q+1)(q^2-1)\big/(q^2-1)=q+1$. 
In other words, when $\beta$ runs over $E^\times$,   
we obtain all $q+1$ simple left ideals $L\beta$ of $M$,
each of them appears $q-1$ times.
\qed

\section{Dihedral group codes}\label{D-codes}

By Theorem~\ref{FG=} and Lemma \ref{M algebra},
from now on we fix the following notation.

\begin{remark}\label{notation C}\rm
$FG=A_0\oplus A_1\oplus\cdots\oplus A_m$, where the ideal
$A_0=FGe_0$ and ideals
$A_t\cong {\rm M}_2(F_t)$, $t=1,\cdots,m$. 
For $t=1,\cdots,m$,  we always assume:
\begin{itemize}
\item[\bf(1)]\vskip-3pt
$Z_t={\rm Z}(A_t)$ which is corresponding to
the center ${\rm Z}\big({\rm M}_2(F_t)\big)$,  so 
$Z_t\cong F_t$ is a field and $\dim_F Z_t=k_t$;
\item[\bf(2)]
fix a field $K_t\subseteq A_t$ which is, by notation of Lemma \ref{M algebra},
 corresponding to the field $E$ contained in ${\rm M}_2(F_t)$ of dimension $2$ 
over $F_t$, in particular, $\dim_F K_t=2k_t$;
\item[\bf(3)]
$C_t$ is the simple left ideal of $A_t$ corresponding to
${\rm M}_2(F_t)\cdot\begin{pmatrix}1&0\\0&0\end{pmatrix}$.
\end{itemize}
And set:
\begin{itemize}
\item[\bf(4)]
$A=A_1\oplus\cdots\oplus A_m$, so $\dim_F A=4k_1+\cdots+4k_m=2(n-1)$;
\item[\bf(5)]
$Z=Z_1\oplus\cdots\oplus Z_m$, so $\dim_F Z=k_1+\cdots+k_m=\frac{n-1}{2}$; 
\item[\bf(6)]
$K=K_1\oplus\cdots\oplus K_m$, so $\dim_F K=2k_1+\cdots+2k_m=n-1$; 
\item[\bf(7)]
$C=C_1\oplus\cdots\oplus C_m$, so $\dim_F C=2k_1+\cdots+2k_m=n-1$;
\item[\bf(8)]
$\widehat C=C_0\oplus C_1\oplus\cdots\oplus C_m$
where $C_0=F\widehat e_0$ as described in Lemma \ref{basic FG}(3),
so $\dim_F \widehat C=1+2k_1+\cdots+2k_m=n$.
\end{itemize}
Then $C$ and $\widehat C$ are dihedral group codes 
of rate $\frac{1}{2}-\frac{1}{2n}$  and $\frac{1}{2}$, respectively.
The multiplicative group of $K$: 
$K^\times=K_1^\times\times\cdots\times K_m^\times$, 
is not a subgroup of the multiplicative group $(FG)^\times$. Let
\begin{equation}\label{K^*}
K^*=\{e_0\}\times K^\times=\{e_0\}\times K_1^\times\times
  \cdots\times K_m^\times,
\end{equation}
where $\{e_0\}$ is the identity subgroup of $A_0^\times$.
Then $K^*$ is a subgroup of $(FG)^\times$.
Note that, if something within $A$, e.g., the code $C$,  is considered, 
then the actions of $K^*$ and $K^\times$ are the same because $e_0C=Ce_0=0$.
\end{remark}

By Theorem \ref{FG=},  for any $j=0,1,\cdots,m$, $\overline 1_{A_j}=1_{A_j}$, 
\begin{equation*}
\overline A_j=\overline{FG\cdot 1_{A_j}}
 =1_{A_j}\cdot FG=FG\cdot 1_{A_j}=A_j,\qquad
 j=0,1,\cdots,m.
\end{equation*}
For any $0\le j\ne j'\le m$, $A_j\overline A_{j'} =A_{j}  A_{j'}=0$.
So, by Lemma \ref{sigma bar}(5),
\begin{equation}\label{ideal orthogonal}
 \big\langle A_j,\,A_{j'} \big\rangle=0, \qquad\forall~0\le j\ne j'\le m.
\end{equation}
  
\begin{lemma}\label{C_t,...}
We keep the notation of Remark \ref{notation C}. Let $1\le t\le m$.

{\bf(1)} 
If $1_{A_t}=e+\overline e$ for a primitive idempotent $e$ of $FH$ with
$e\ne\overline e$, then $C_t\overline C_t=0$ hence 
$\langle C_t, C_t\rangle=0$.

{\bf(2)} 
Assume that $1_{A_t}=e$ for a primitive idempotent $e$ of $FH$ with
$e=\overline e$. 
 
~ {\bf(i)} If ${\rm char}\,F=2$, then $C_t\overline C_t=0$ 
  hence $ \langle C_t, C_t\rangle=0$.
  
~  {\bf(ii)} If ${\rm char}\,F$ is odd, then $C_t\overline C_t\ne 0$ hence 
   $ \langle C_t, C_t\rangle\ne 0$.
\end{lemma}

{\it Proof}. (1).~ By Lemma \ref{FHe} and its 
isomorphism Eq.\eqref{FHe=M_2}, 
$e$ is corresponding to $\begin{pmatrix}1&0\\ 0&0\end{pmatrix}$,
so $C_t=A_t e$.
Then $C_t\overline C_t=A_t e\overline eA_t=0$, since $e\overline e=0$.

(2).~ By Eq.\eqref{eta nu}, $\varepsilon-\nu
=\begin{pmatrix}2&0\\ g&0 \end{pmatrix}$, 
whose first column $\begin{pmatrix}2\\ g \end{pmatrix}\ne 0$, 
see the note before Eq.\eqref{eta nu}. 
So ${\rm M}_2(F_t)\begin{pmatrix}1&0\\0&0 \end{pmatrix}=
{\rm M}_2(F_t)\begin{pmatrix}2&0\\g&0 \end{pmatrix}$.
By the isomorphism Eq.\eqref{FGe cong},
$e$ and $ve$ are corresponding to $\varepsilon$ and $\nu$ respectively.
So $C_t=A_t(e-ve)$. By the definition of the ``bar'' map in Eq.\eqref{bar map},
$\overline  v=v$.  So 
\begin{equation*}
C_t\overline C_t=A_t(e-ve)(e-ve)A_t=A_t(2e-2ve)A_t.
\end{equation*}
If ${\rm char}\,F=2$, then $2e-2ve=0$, hence $C_t\overline C_t=0$.
Thus (i) holds.
If  ${\rm char}\,F$ is odd, then $2e\ne 0$. 
Note that $2e\in FH$ , $2ve\in vFH$;
by Lemma \ref{basic FG}, $2e-2ve\ne 0$. Hence $C_t\overline C_t\ne 0$.
\qed

\begin{theorem}\label{char 2}
If ${\rm char}\,F=2$, then for any $\beta\in K^*$,
$\widehat C\beta$ is a self-dual dihedral group code.
\end{theorem}

{\it Proof}.~
First we show that $\widehat C$ is self-dual.
For any $c=c_0+c_1+\cdots+c_m$ and $c'=c_0'+c_1'+\cdots+c_m'$, 
where $c_j,c_j'\in C_j$, $j=0,1,\cdots,m$, by Eq.\eqref{ideal orthogonal},
$$
\langle c_0+c_1+\cdots+c_m, c_0'+c_1'+\cdots+c_m'\rangle
=\langle c_0,c_0'\rangle+\langle c_1,c_1'\rangle+\cdots+\langle c_m,c_m'\rangle.
$$
By Lemma \ref{sigma bar}(2) and Lemma \ref{basic FG}(3),
$\langle\widehat e_0, \widehat e_0\rangle
=\sigma(\widehat e_0\overline{\widehat e}_0)=\sigma(2\widehat e_0)
=\frac{2}{n}=0$,
hence $\langle c_0,c_0'\rangle=0$.
By Lemma \ref{C_t,...},  
$\langle c_t,c_t'\rangle=0$ for $1\le t\le m$.
That is, $\langle c, c'\rangle=0$.
So $\widehat C$ is self-orthogonal.
Further, the rate ${\rm R}(\widehat C)=\frac{1}{2}$.
Thus $\widehat C$ is self-dual.

For the general case,  since $\overline{\widehat C}\widehat C=0$ 
(see Lemma \ref{sigma bar}(5)), we have
$$
\overline{\widehat C\beta}\cdot\widehat C\beta
 =\overline \beta \overline{\widehat C}\widehat C\beta
=0, \quad\mbox{hence}~~
\big\langle \widehat C\beta,\widehat C\beta\big\rangle=0.
$$
And ${\rm R}(\widehat C\beta)=\frac{1}{2}$. We obtain that
$\widehat C\beta$ is self-dual.
\qed

\medskip
The word ``module'' in this paper 
means a left module, except for other declarations. 
Note that $\widehat C$ is an $FG$-module.

\begin{lemma}\label{submodule}
If $D$ is an $FG$-submodule of $\widehat C$, then 
$$
 D=(D\cap C_{0})\oplus(D\cap C_{1})\oplus \cdots\oplus
 (D\cap C_{m}),
$$
and each $D\cap C_j$ is either $0$ or $C_j$, for $j=0,1,\cdots,m$.
\end{lemma} 

{\it Proof}.~ 
The identity 
$1_{FG}=e_0+1_{A_1}+\cdots+1_{A_m}$
is a sum of central idempotents, and  $e_01_{A_t}=0$, 
$1_{A_t}1_{A_{t'}}=0$ for $1\le t\ne t'\le m$. 
For any $d\in D$ we have
$$
d=(e_0+1_{A_1}+\cdots+1_{A_m})d
=e_0d+ 1_{A_1}d+\cdots+ 1_{A_m}d
\in (D\cap C_{0}) 
\oplus \cdots\oplus (D\cap C_{m}).
$$
So the equality of the lemma holds. Since the $FG$-module $C_j$ is simple, 
$D\cap C_j$ is either $0$ or $C_j$.
\qed

\begin{theorem}\label{char odd}
Assume that ${\rm char}\,F$ is odd, and $\beta\in K^*$.

{\bf(1)}
If ${\rm ord}_{{\Bbb Z}_n^\times}\!(q)\!$ is odd, then $C\beta$ is a maximal 
self-orthogonal code of rate $\frac{1}{2}-\frac{1}{2n}$.

{\bf(2)}
If $-1\in\langle q\rangle_{{\Bbb Z}_n^\times}$, then $\widehat C\beta$ 
is an LCD code of rate $\frac{1}{2}$.
\end{theorem}

{\it Proof}. 
(1).~ By Lemma~\ref{n coprime to q}(2) and Lemma~\ref{C_t,...}(1),
$C_t\overline C_t=0$ for $t=1,\cdots,m$. 
By the same argument as in the proof of Theorem \ref{char 2},
$C\beta$ is a self-orthogonal code of rate $\frac{1}{2}-\frac{1}{2n}$.
But this  time $\langle \widehat e_0,\widehat e_0\rangle=\frac{2}{n}\ne 0$,
$C_0=F\widehat e_0$ is not self-orthogonal, hence $C$ is maximal
self-orthogonal.

(2).~ Write $\beta=e_0+\beta_1+\cdots+\beta_m$, 
where $\beta_t\in K_t^\times$
for $t=1,\cdots,m$. Then
$$ \widehat C\beta
 =C_0\oplus C_1\beta_1\oplus\cdots\oplus  C_m\beta_m.
$$
As shown above, $C_0$ is not self-orthogonal.
For $1\le t\le m$,
by Lemma~\ref{n coprime to q}(3) 
and Lemma~\ref{C_t,...}(2), $C_t\overline C_t\ne 0$ 
(i.e. $\overline C_t C_t\ne 0$); hence
$$
\overline{C_t\beta_t}\cdot C_t\beta_t
=\overline \beta_t \overline C_t C_t\beta\ne 0,
$$
i.e., $C_t\beta_t$ is not self-orthogonal.
Denote $D=(\widehat C\beta)\bigcap(\widehat C\beta)^\bot$.
By Lemma~\ref{submodule},
$D=(D\cap C_0)\oplus \bigoplus\limits_{t=1}^m (D\cap  C_t\beta_t$).
But, $D\cap C_0$, $D\cap  C_t\beta_t$ must be self-orthogonal, hence
$D\cap C_0\ne C_0$, $D\cap  C_t\beta_t\ne C_t\beta_t$.
By Lemma~\ref{submodule}, 
$D\cap C_0=0$, $D\cap  C_t\beta_t=0$.
Then $D=0$, and $\widehat C\beta$ is an LCD code. 
\qed

\section{Random dihedral codes of rate $\frac{1}{2}-\frac{1}{2n}$}
\label{random D-codes}

Keep the assumptions in Remark \ref{notation G} and Remark \ref{notation C}.
For $k_t$ in Remark \ref{notation C}(1),
we further assume that $k_1\le k_2\le\cdots\le k_m$.
By Corollary~\ref{k_1 lambda}, $2k_1\ge\lambda(n)$.
By Lemma~\ref{n coprime to q}(1) and Lemma~\ref{good primes}, 
we can further assume $\lambda(n)>\log_q n$. 
Hence, 
in the following we always assume that 
\begin{equation}\label{k_1<...}\textstyle
\frac{1}{2}\log_qn< \frac{1}{2}\lambda(n)\le k_1\le k_2\le\cdots\le k_m.
\end{equation}

From now on, let $\delta$ be a real number satisfying that
\begin{equation}\label{delta,n}\textstyle
 0<\delta<1-q^{-1},\qquad \frac{1}{4}-h_q(\delta)
  -\frac{\log_q n}{\lambda(n)}>0.
\end{equation}
Note that, if $h_q(\delta)<\frac{1}{4}$,
by Lemma~\ref{n coprime to q}(1) and Lemma~\ref{good primes}, there are
infinitely many odd integers $n>1$ coprime to $q$ such that 
$\frac{1}{4}-h_q(\delta)-\frac{\log_q n}{\lambda(n)}$ 
are positively bounded from below. 

For any left ideal $L$ of $FG$ and any $(\alpha,\beta)\in K^*\times K^*$,
 $\alpha$ is a unit of~$FG$, see Eq.\eqref{K^*}; so  
 $(FG)\alpha=FG=\alpha(FG)$, hence 
$\alpha L\beta=\alpha\cdot FG\cdot L\beta=FG\cdot L\beta= L\beta$
 is a left ideal.

\begin{definition}\label{random C}\rm
Consider $K^*\times K^*$ as a probability space
with equal probability for every sample.
Let $(\alpha,\beta)\in K^*\times K^*$. We have the following:

{\bf(1)}
$C_{\alpha,\beta}:=\alpha C\beta=C\beta$  
is a random $FG$-code 
with rate $R(C_{\alpha,\beta})=\frac{1}{2}-\frac{1}{2n}$.

{\bf(2)}
$\Delta(C_{\alpha,\beta})=\frac{{\rm w}(\alpha C\beta)}{2n}$ 
is a random variable.

{\bf(3)}
For $c\in C$, define a $0$-$1$ variable:
$X_c=\begin{cases}
  1, & 0<\frac{{\rm w}(\alpha c\beta)}{2n}\le \delta;\\[2pt]
  0, & \mbox{otherwise}. \end{cases}$
 
{\bf(4)} Let $X=\sum_{c\in C}X_c$, which stands for 
the number of the non-zero codewords~$\alpha c\beta$  
whose relative weights are at most $\delta$. 
\end{definition}

By $\Pr\big(\Delta(C_{\alpha,\beta})\le\delta\big)$ we denote the probability
that $\Delta(C_{\alpha,\beta})\le\delta$,
 and by ${\rm E}(X)$ we denote the expectation of the random variable $X$.
Then
$$
 \Pr\big(\Delta(C_{\alpha,\beta})\le\delta\big)=\Pr(X\ge 1).
$$ 
By  Markov's inequality (c.f. \cite[Theorem 3.1]{MU}),
for the non-negative integer variable $X$ we have
$\Pr(X\ge 1)\le{\rm E}(X)$.  So
\begin{equation}\label{<E(X)}
\Pr\big(\Delta(C_{\alpha,\beta})\le\delta\big)\le{\rm E}(X).
\end{equation}
If $c=0$ then $X_0=0$ obviously.
By the linearity of expectations,
\begin{equation}\label{E(X)=}\textstyle
{\rm E}(X)=\sum_{c\in C}{\rm E}(X_c)
  =\sum_{0\ne c\in C}{\rm E}(X_c).
\end{equation}
Since $X_c$ is a 0-1 variable,
\begin{equation}\label{E(X_c)}\textstyle
{\rm E}(X_c)=\Pr(X_c=1)=\Pr\big(0<\frac{{\rm w}(\alpha c\beta)}{2n}\le \delta\big).
\end{equation}

We estimate ${\rm E}(X_c)$ for $0\ne c\in C$.
Set $C_t^+=C_t\backslash\{0\}$,  $t=1,\cdots,m$.
For the non-zero $c\in C$, there is a subset 
$\omega=\{t_1,\cdots,t_r\}\subseteq \{1,2,\cdots,m\}$ 
such that 
\begin{equation}\label{c=c_t_1+...}
 c=c_{t_1}+c_{t_2}+\cdots+c_{t_r}, ~~~
  c_{t_j}\in C_{t_j}^+=C_{t_j}\backslash\{0\}, ~~ j=1,\cdots,r.
\end{equation}
Then
$
 Ac=A_{t_1}c_{t_1}\oplus\cdots\oplus A_{t_r}c_{t_r}. 
$
Since $A_{t_j}c_{t_j}\ne 0$ is a submodule of $C_{t_j}$ and $C_{t_j}$ is simple,
we have $A_{t_j} c_{t_j}=C_{t_j}$. So,
\begin{equation}\label{c in Ac}\textstyle
Ac=C_{t_1}\oplus\cdots\oplus C_{t_r}, \quad
 \end{equation}
 and $\dim_F(Ac)=2k_{t_1}+\cdots+2k_{t_r}$ is even.
Denote 
\begin{equation}\label{l_c}
\textstyle\ell_c=\frac{\dim_F(Ac)}{2}=k_{t_1}+\cdots+k_{t_r},
\end{equation}
then $k_1\le\ell_c\le\frac{n-1}{2}$
(cf.
Eq.\eqref{k_1<...} and Remark \ref{notation C}(5)).

\begin{lemma}\label{X_c}
Let the notation be as above. Then
$
{\rm E}(X_c) < q^{-3\ell_c+4\ell_c h_q(\delta)+4}.
$
\end{lemma}

{\it Proof}. 
Let $\widetilde\omega=\{1,2,\cdots,m\}\backslash\omega
=\{1,2,\cdots,m\}\backslash\{t_1,t_2,\cdots,t_r\}$. 
Let
$$\begin{array}{l}
A_\omega=A_{t_1}\oplus\cdots\oplus A_{t_r}, \quad 
 A_{\widetilde\omega}=\textstyle\bigoplus\limits_{t\in\widetilde\omega}A_t,
 \quad\mbox{hence}\quad A=A_\omega\oplus A_{\widetilde\omega};\\[8pt]
K_{\omega}^\times=K_{t_1}^\times \times \cdots\times K_{t_r}^\times,
 \quad K_{\widetilde\omega}^\times
   =\mathop{\times}\limits_{t\in\widetilde\omega}K_t^\times,
 \quad\mbox{hence}\quad 
 K^\times=K_{\omega}^\times \times K_{\widetilde\omega}^\times;\\[6pt]
Z_{\omega}^\times=Z_{t_1}^\times 
  \times \cdots\times Z_{t_r}^\times.
\end{array}
$$
For $(\alpha,\beta), (\alpha',\beta')\in K^* \times K^*$, 
by Eq.\eqref{K^*}, we can write 
$\alpha=e_0+\alpha_\omega+\alpha_{\widetilde\omega}$
with 
$\alpha_{\omega}\in K_{\omega}^\times$ and 
$\alpha_{\widetilde\omega}\in K_{\widetilde\omega}^\times$; 
since $e_0c=0=ce_0$,
$$\alpha c\beta
= (\alpha_\omega+\alpha_{\widetilde\omega})c(\beta_\omega+\beta_{\widetilde\omega})
=\alpha_\omega c\beta_\omega,~~\mbox{and}~~
\alpha' c\beta'=\alpha'_\omega c\beta'_\omega.
$$
By Lemma \ref{M algebra}(2),
$\alpha c\beta=\alpha'c\beta'$
if and only if 
$\alpha'^{-1}_\omega\alpha_\omega=\beta'_\omega\beta^{-1}_\omega\in  Z_{\omega}^\times$, if and only if
there are $z_{\omega}\in Z_{\omega}^\times$ and 
$(\alpha'_{\widetilde\omega}, \beta'_{\widetilde\omega})\in K_{\widetilde\omega}^\times \times K_{\widetilde\omega}^\times$
such that 
$$
  \alpha'
  =\alpha_{\omega}z_{\omega}^{-1}+ \alpha'_{\widetilde\omega}, 
  \quad
  \beta'=\beta_{\omega}z_{\omega}+\beta'_{\widetilde\omega}.
$$
So, for $d\in K^* cK^*$, 
there are exactly 
$\big|Z_{\omega}^\times\big|\cdot \big|K_{\widetilde\omega}^\times \big|^2$ 
paires $(\alpha,\beta)$ in 
$K^* \times K^*$ such that $\alpha c\beta=d$.
Since
$$K^* cK^*=K^\times cK^\times
=K_{t_1}^\times c_{t_1}K_{t_1}^\times 
 \times \cdots \times K_{t_r}^\times c_{t_r}K_{t_r}^\times
 \subseteq A_{t_1}\oplus\cdots\oplus A_{t_r}=A_{\omega},
$$
and $A_\omega$ is an ideal in $FG$ of dimension $4\ell_c$ over $F$,
we get
$$
\big|(K^* cK^*)^{\le\delta}\big|
\le\big|(A_\omega)^{\le\delta}\big|\le q^{4\ell_c h_q(\delta)},
$$
where the last inequality follows from Lemma \ref{at most delta}.
Thus, there are at most
 $\big|Z_{\omega}^\times\big|\cdot \big|K_{\widetilde\omega}^\times \big|^2
 \cdot q^{4\ell_c h_q(\delta)}$
pairs $(\alpha,\beta)\in K^* \times K^*$ such that
$0<\frac{{\rm w}(\alpha c\beta)}{2n}\le\delta$. 
By Eq.\eqref{E(X_c)}, 
 we obtain that
$$
{\rm E}(X_c)\le
\frac{\big|Z_{\omega}^\times\big|\cdot 
 \big|K_{\widetilde\omega}^\times \big|^2\cdot q^{4\ell_c h_q(\delta)}}
 {|K^* \times K^*|}
 =\frac{\big|Z_{\omega}^\times\big|\cdot q^{4\ell_c h_q(\delta)}}
 {|K_\omega^\times|^2}.
$$
We estimate the cardinalities 
$|Z_{\omega}^\times|$ and $|K_\omega^\times|$ as follows:
$$\begin{array}{c}
 |Z_{\omega}^\times|=\prod\limits_{j=1}^r(q^{k_{t_j}}-1)
 <q^{k_{t_1}}\cdot\cdots\cdot q^{k_{t_r}}
 =q^{k_{t_1}+\cdots+k_{t_r}}
 =q^{\ell_c},\\
|K_{\omega}^\times|=\prod\limits_{j=1}^r(q^{2k_{t_j}}-1)
>q^{2(k_{t_1}+\cdots+k_{t_r})-2}
=q^{2\ell_c-2};
\end{array}$$
where the second inequality follows from Lemma \ref{k_t} and Eq.\eqref{k_1<...}.
Then 
$$
{\rm E}(X_c)
\le
\frac{q^{\ell_c} \cdot q^{4\ell_c h_q(\delta)}}
 {(q^{2\ell_c-2})^2}
 = q^{-3\ell_c+4\ell_c h_q(\delta)+4}.
\eqno\qed $$

By Lemma \ref{submodule}, the following $\Omega$ is
the set of all $A$-submodules of $C$:
\begin{equation}\label{Eq Omega}
 \Omega=\big\{C_{t_1}\oplus\cdots\oplus C_{t_r}~\big|~\{t_1,\cdots,t_r\}\subseteq  \{1,2,\cdots,m\}\big\}.
\end{equation}
\begin{lemma}\label{Omega}
For $D=C_{t_1}\oplus\cdots\oplus C_{t_r}\in\Omega$,
let $D^+=C_{t_1}^+\oplus\cdots\oplus C_{t_r}^+$
where $C_{t_j}^+=C_{t_j}\backslash\{0\}$ as before.
For $k_1\le\ell\le\frac{n-1}{2}$,
let $\Omega_\ell=\{D\in\Omega\,|\,\dim_FD=2\ell\}$ 
 (it is possible that $\Omega_\ell=\emptyset$).
 Then
 
 {\bf(1)}~ $\Omega=\bigcup\limits_{\ell=k_1}^{(n-1)/2}\Omega_\ell$,
and $C\backslash\{0\}\label{C-0}
=\bigcup\limits_{\ell=k_1}^{(n-1)/2}\bigcup\limits_{D\in\Omega_\ell} D^+$.

{\bf(2)}~ $|\Omega_\ell|\le n^{\ell/k_1}$.
\end{lemma}

{\it Proof}.~ 
(1) is proved directly.

If $C_{t_1}\oplus\cdots\oplus C_{t_r}\in\Omega_\ell$,
then $k_{t_1}+\cdots+k_{t_r}=\ell$; in particular, $r\le\ell/k_1$.
Thus
$$\textstyle
 |\Omega_\ell|\le\sum_{j=1}^{\ell/k_1}\binom{m}{j}
\le\sum_{j=1}^{\ell/k_1}\binom{n}{j}\le n^{\ell/k_1}.
\eqno\qed
$$

\begin{theorem}\label{E(X)<}
${\rm E}(X) <
 q^{-4k_1\big(\frac{1}{4}-h_q(\delta)-\frac{\log_q n}{2k_1}\big)+4}.
$
\end{theorem}

{\it Proof}.~ 
By Eq.\eqref{E(X)=} and Lemma \ref{Omega}(1), we have 
$$ \textstyle
 {\rm E}(X)
 =\sum\limits_{0\ne c\in C}{\rm E}(X_c)
 =\sum\limits_{\ell=k_1}^{(n-1)/2}\sum\limits_{D\in\Omega_\ell}
 \sum\limits_{c\in D^+}{\rm E}(X_c).
$$
For $D\in\Omega_\ell$ and $c\in D^+$, 
$\ell_c=\frac{1}{2}\dim_FD=\ell$, see Eq.\eqref{c in Ac} and Eq.\eqref{l_c}. 
By Lemma \ref{X_c} and Lemma \ref{Omega}(2),
\begin{align*}
&\textstyle\sum\limits_{D\in\Omega_\ell}
 \sum\limits_{c\in D^+}{\rm E}(X_c)<
 \sum\limits_{D\in\Omega_\ell}\sum\limits_{c\in D^+}
  q^{-3\ell+4\ell h_q(\delta)+4}\\
 &\textstyle< \sum\limits_{D\in\Omega_\ell}
 q^{2\ell}\cdot q^{-3\ell+4\ell h_q(\delta)+4}
\le  n^{\frac{\ell}{k_1}}
  q^{-\ell+4\ell h_q(\delta)+4}\\
&= q^{-4\ell\big(\frac{1}{4}-h_q(\delta)-\frac{\log_q n}{4k_1}\big)+4}
\le q^{-4k_1\big(\frac{1}{4}-h_q(\delta)-\frac{\log_q n}{4k_1}\big)+4}.
\end{align*}
The last inequality holds since $\ell\ge k_1$ (by Eq.\eqref{l_c}) and 
$\frac{1}{4}-h_q(\delta)-\frac{\log_q n}{4k_1}>0$ (by Eq.\eqref{delta,n}).
Further, $\frac{n-1}{2}-k_1+1\le n=q^{\log_q n}$. So
\begin{align*}
{\rm E}(X)
&\textstyle < \sum\limits_{\ell=k_1}^{(n-1)/2} 
 q^{-4k_1\big(\frac{1}{4}-h_q(\delta)-\frac{\log_q n}{4k_1}\big)+4}\\
 &\textstyle =\big(\frac{n-1}{2}-k_1+1\big)
 q^{-4k_1\big(\frac{1}{4}-h_q(\delta)-\frac{\log_q n}{4k_1}\big)+4}\\
 &\le q^{\log_q n}
 q^{-4k_1\big(\frac{1}{4}-h_q(\delta)-\frac{\log_q n}{4k_1}\big)+4}.
\end{align*}
That is, 
${\rm E}(X) <
 q^{-4k_1\big(\frac{1}{4}-h_q(\delta)-\frac{\log_q n}{2k_1}\big)+4}$. 
\qed

\begin{theorem}\label{thm for C}
$\Pr\big(\Delta(C_{\alpha,\beta})\le\delta\big)<
 q^{-2\lambda(n)\big(\frac{1}{4}
      -h_q(\delta)-\frac{\log_q n}{\lambda(n)}\big)+4}.$
\end{theorem}

{\it Proof}. 
Combining Theorem~\ref{E(X)<} with Eq.\eqref{<E(X)}, we get 
$$\Pr\big(\Delta(C_{\alpha,\beta})\le\delta\big)<
 q^{-4k_1\big(\frac{1}{4}-h_q(\delta)-\frac{\log_q n}{2k_1}\big)+4}.
$$
By Corollary \ref{k_1 lambda}, $2k_1\ge\lambda(n)$; 
and by Eq.\eqref{delta,n}, 
$$\textstyle
 \frac{1}{4}-h_q(\delta)-\frac{\log_q n}{2k_1}>
 \frac{1}{4}-h_q(\delta)-\frac{\log_q n}{\lambda(n)}>0.
$$
we get that 
$$
q^{-4k_1\big(\frac{1}{4}-h_q(\delta)-\frac{\log_q n}{2k_1}\big)+4}
\le
q^{-2\lambda(n)\big(\frac{1}{4}
      -h_q(\delta)-\frac{\log_q n}{\lambda(n)}\big)+4}.
      \eqno\qed
$$

\section{Random dihedral codes of rate $\frac{1}{2}$}
\label{random D-codes 1/2}

Keep the notation in \S\ref{random D-codes}. In particular,
Eq.\eqref{k_1<...}, Eq.\eqref{delta,n} hold and $K^* \times K^*$ is considered as a 
probability space with equal probability for each sample.
We start from $\widehat C=C_0\oplus C$ where $C_0=F\widehat e_0$, 
see Remark \ref{notation C}(8). Then
\begin{equation}\label{random hat C} 
 \widehat C_{\alpha,\beta}=\alpha\widehat C\beta,\qquad 
  (\alpha,\beta)\in K^* \times K^*,
\end{equation}
is a random code with $R(\widehat C_{\alpha,\beta})=\frac{1}{2}$.
Define 
$$\textstyle
  \widehat X_c =\begin{cases}1, & 
  0<\frac{{\rm w}(\alpha c\beta)}{2n}\le \delta;\\[2pt]
  0, & \mbox{otherwise}; \end{cases} \quad c\in\widehat C;
  \qquad\mbox{and}~~
 \widehat X=\sum\limits_{c\in\widehat C}\widehat X_c.
$$
We still have
\begin{equation}
 \Pr\big(\Delta(\widehat C_{\alpha,\beta})\le\delta\big)
  =\Pr(\widehat X\ge 1)\le {\rm E}(\widehat X).
\end{equation}

Recall that 
$\Omega=\bigcup_{\ell=k_1}^{(n-1)/2}\Omega_\ell$
is the set of all non-zero submodules of $C$, 
see Eq.\eqref{Eq Omega} and Lemma \ref{Omega}. 
It is easy to check the following.

\begin{lemma}\label{C_0 Omega}
Denote $C_0\oplus\Omega=\{C_0\oplus D\mid D\in\Omega\}$.
For $D'=C_0\oplus D$ with $D\in\Omega$,
let $ D'^+=C_0^+\oplus D^+$ ($D^+$ is defined in Lemma \ref{Omega}). Then
$$\textstyle
\widehat C\backslash\{0\}
 =C_0^+
 \bigcup\, (\bigcup\limits_{D\in\Omega}D^+)
 \bigcup\,(\bigcup\limits_{D'\in C_0\oplus \Omega}D'^+).
$$
\end{lemma}

We already have the estimation of 
$\sum\limits_{D\in\Omega}\sum\limits_{c\in D^+}{\rm E}(\widehat X_c)$, 
see Theorem \ref{E(X)<}.

For $0\ne c\in C_0$ and 
$(\alpha,\beta)\in K^* \times K^*$,
it is trivial that $\alpha c\beta=c$ and $\frac{{\rm w}(\alpha c\beta)}{2n}=1$. 
So ${\rm E}(\widehat X_c)=0$. Hence
\begin{equation}\label{C_0*}\textstyle
 \sum_{c\in C_0^+}{\rm E}(\widehat X_c) =0.
\end{equation}

\begin{lemma} \label{hat Omega}
$\sum\limits_{D'\in(C_0\oplus\Omega	)}
 \sum\limits_{c\in D'^+}{\rm E}(\widehat X_c)< q^2
 q^{-4k_1\big(\frac{1}{4}-h_q(\delta)-\frac{\log_q n}{2k_1}\big)+4}$. 
\end{lemma}

{\it Proof}.~ 
For $k_1\le \ell\le\frac{n-1}{2}$,
let $D'=C_0\oplus D$ with $D\in\Omega_\ell$, 
and let $c'\in D'^+$. 
Similarly to the proof of Lemma \ref{X_c}, 
we assume that $\omega=\{t_1,\cdots,t_r\}
\subseteq\{1,\cdots,m\}$ such that
$$ c'=c_0+c_{t_1}+\cdots+c_{t_r},\quad 
 c_0\in C_0^+,~~ c_{t_j}\in C_{t_j}^+,~ j=1,\cdots,r;
$$  
and construct 
$$A'_{\omega}=C_0\oplus A_{t_1}\oplus\cdots\oplus A_{t_r},
\quad 
 K_{\omega}'^\times=\{e_0\}\times K_{t_1}^\times 
  \times\cdots\times K_{t_r}^\times. $$
It is the same as in the proof of Lemma \ref{X_c},
except that $\dim_F(A'_{\omega})=4\ell+1$, hence
 $$
\big|(K^* c'K^*)^{\le\delta}\big|
\le\big|(A'_\omega)^{\le\delta}\big|\le q^{(4\ell+1)h_q(\delta)}.
$$
We obtain
\begin{equation*}
 {\rm E}(\widehat X_c)
 <q^{-3\ell+4\ell h_q(\delta)+h_q(\delta) + 4}
 <q^{-3\ell+4\ell h_q(\delta)+5}.
\end{equation*}
Because $| D'^+|<|D'|=q^{2\ell+1}$, 
$$\textstyle
\sum\limits_{c\in D'^+}{\rm E}(\widehat X_c)
< q^{2\ell+1}q^{-3\ell+4\ell h_q(\delta) + 5}
= q^{-\ell+4\ell h_q(\delta) + 6}.
$$
Then, similarly to Theorem~\ref{E(X)<}, we obtain
$$\begin{array}{rcl}
\sum\limits_{D'\in(C_0\oplus\Omega)}
 \sum\limits_{c\in D'^+}{\rm E}(\widehat X_c)
&=& 
\sum\limits_{\ell=k_1}^{(n-1)/2}
\sum\limits_{ D'\in (C_0\oplus\Omega_\ell)}
\sum\limits_{c\in D'^+}{\rm E}(\widehat X_c)\\[5mm]
&<& 
q^{-4k_1\big(\frac{1}{4}-h_q(\delta)-\frac{\log_q n}{2k_1}\big)+6}.
\end{array}
$$
That is,
$$\textstyle
\sum\limits_{D'\in(C_0\oplus\Omega)}
 \sum\limits_{c\in D'^+}{\rm E}(\widehat X_c)
 < q^2
 q^{-4k_1\big(\frac{1}{4}-h_q(\delta)-\frac{\log_q n}{2k_1}\big)+4} .
 \eqno\qed
$$

\begin{theorem}\label{thm for widehat C}
 ${\rm E}(\widehat X)< (1+q^2)
 q^{-4k_1\big(\frac{1}{4}-h_q(\delta)-\frac{\log_q n}{2k_1}\big)+4}$.
\end{theorem}

{\it Proof}.~ By Lemma \ref{C_0 Omega}, Eq.\eqref{C_0*}, 
Theorem \ref{E(X)<} and Lemma \ref{hat Omega},
$$\begin{array}{rl}
{\rm E}(\widehat X)\kern-4pt
 &=\sum\limits_{0\ne c\in\widehat C}{\rm E}(\widehat X_c)\\[5mm]
 &=\sum\limits_{c\in C_0^+}{\rm E}(\widehat X_c)
   + \sum\limits_{D\in\Omega}\sum\limits_{c\in D^+}{\rm E}(\widehat X_c)
   + \sum\limits_{D'\in(C_0\oplus\Omega)}
    \sum\limits_{c\in D'^+}{\rm E}(\widehat X_c)\\[5mm]
  & <~ 0 + 
 q^{-4k_1\big(\frac{1}{4}-h_q(\delta)-\frac{\log_q n}{2k_1}\big)+4}
 + q^2
 q^{-4k_1\big(\frac{1}{4}-h_q(\delta)-\frac{\log_q n}{2k_1}\big)+4}
 \\[3mm]
 &=~(1+q^2)
 q^{-4k_1\big(\frac{1}{4}-h_q(\delta)-\frac{\log_q n}{2k_1}\big)+4}.
\end{array}
$$
We are done. \qed

\medskip
Similarly to Theorem~\ref{thm for C}, we obtain:

\begin{theorem}\label{thm for hat C}~
$\Pr\big(\Delta(\widehat C_{\alpha,\beta})\le\delta\big) 
< (1+q^2) q^{-2\lambda(n)\big(\frac{1}{4}-h_q(\delta)
   -\frac{\log_q n}{\lambda(n)}\big)+4}.$
\end{theorem}

\section{Proofs of the main theorems }\label{proof of thms}

For a sequence $n_1,n_2,\cdots$ of odd positive integers $n_i$ coprime to $q$
with $n_i\to\infty$, we have a sequence 
$G^{(1)},G^{(2)},\cdots$ of dihedral groups $G^{(i)}$ of order $2n_i$,
and have random $FG^{(i)}$-codes:

$\bullet$~ $C_{\alpha,\beta}^{(i)}$
~of rate $\frac{1}{2}-\frac{1}{2n_i}$, defined in Definition~\ref{random C};

$\bullet$~
$\widehat C_{\alpha,\beta}^{(i)}$
~of rate $\frac{1}{2}$, defined in Eq.\eqref{random hat C};

\noindent hence we have two sequences of random dihedral codes:
\begin{eqnarray}
\label{Cs}
 C_{\alpha,\beta}^{(1)},~ C_{\alpha,\beta}^{(2)},~ 
  C_{\alpha,\beta}^{(3)},~\cdots; 
\\[1mm] \label{hat Cs}
\widehat C^{(1)}_{\alpha,\beta},~ 
 \widehat C_{\alpha,\beta}^{(2)},~ 
  \widehat C_{\alpha,\beta}^{(3)},~ \cdots. 
\end{eqnarray}

\begin{theorem}\label{thm1'}
Assume that $0<\delta<1-q^{-1}$ and $0<h_q(\delta)<\frac{1}{4}$.
Assume that ${\rm char}\,F=2$. Then there is a sequence 
$n_1,n_2,\cdots$ of odd integers $n_i$ coprime to $q$ with $n_i\to\infty$
such that

{\bf(1)} 
The sequence in Eq.\eqref{hat Cs} consists of self-dual dihedral codes;

{\bf(2)} 
$\lim\limits_{i\to\infty}
\Pr\big(\Delta(\widehat C_{\alpha,\beta}^{(i)})>\delta\big)=1$.
\end{theorem}

{\it Proof}.~
By Lemma~\ref{good primes},
there is a series $n_1,n_2,\cdots$ of odd integers coprime to $q$
such that $\lim\limits_{i\to\infty}\frac{\log_q n_i}{\lambda(n_i)}=0$.
Then (1) follows from Theorem~\ref{char 2}. 
And, since $\frac{1}{4}-h_q(\delta)-\frac{\log_q n_i}{\lambda(n_i)}>0$
and $\lambda(n_i)\to\infty$, by Theorem~\ref{thm for hat C},
$$
\lim\limits_{i\to\infty}
\Pr\big(\Delta(\widehat C_{\alpha,\beta}^{(i)})\le\delta\big)
< \lim\limits_{i\to\infty}
 (1+q^2) q^{-2\lambda(n_i)\big(\frac{1}{4}-h_q(\delta)
   -\frac{\log_q n_i}{\lambda(n_i)}\big)+4}
=0. 
$$
That is, (2) holds.
\qed

\medskip
Theorem~\ref{thm1} is obviously a consequence of  
Theorem~\ref{thm1'}. On the other hand, 
Theorem~\ref{thm2} is a consequence of the following theorem.

\begin{theorem}\label{}
Assume that $0<\delta<1-q^{-1}$ and $0<h_q(\delta)<\frac{1}{4}$.
Assume that ${\rm char}\,F$ is odd.

{\bf(1)} 
There is a sequence 
$n_1,n_2,\cdots$ of odd integers $n_i$ coprime to $q$ with $n_i\to\infty$
such that Eq.\eqref{Cs} is a sequence of maximal self-orthogonal
dihedral codes of rate $\frac{1}{2}-\frac{1}{2n_i}$ and
$\lim\limits_{i\to\infty}
\Pr\big(\Delta(C_{\alpha,\beta}^{(i)})>\delta\big)=1$.

{\bf(2)} 
There is a sequence $n_1,n_2,\cdots$ of odd integers 
$n_i$ coprime to $q$ with ${n_i\to\infty}$ such that 
Eq.\eqref{hat Cs} is a sequence of LCD dihedral codes of rate $\frac{1}{2}$ 
and
$\lim\limits_{i\to\infty}
\Pr\big(\Delta(\widehat C_{\alpha,\beta}^{(i)})>\delta\big)=1$.
\end{theorem} 

{\it Proof}.
(1). By Corollary~\ref{infinitely n_i}(1), there is a sequence 
$n_1,n_2,\cdots$  of odd integers $n_i$ coprime to $q$ such that
${\rm ord}_{{\Bbb Z}_{n_i}^\times}(q)$ are all odd
and $\lim\limits_{i\to\infty}\frac{\log_q n_i}{\lambda(n_i)}=0$.
By Theorem~\ref{char odd}(1),
Eq.\eqref{Cs} is a sequence of maximal self-orthogonal
dihedral codes of rate $\frac{1}{2}-\frac{1}{2n_i}$. 
By Theorem~\ref{thm for C},
$$
\lim\limits_{i\to\infty}
\Pr\big(\Delta(C_{\alpha,\beta}^{(i)})\le\delta\big)
< \lim\limits_{i\to\infty}
  q^{-2\lambda(n_i)\big(\frac{1}{4}-h_q(\delta)
   -\frac{\log_q n_i}{\lambda(n_i)}\big)+4}
=0. 
$$

(2). The proof is similar to the above by citing 
Corollary~\ref{infinitely n_i}(2), Theorem~\ref{char odd}(2)
and Theorem~\ref{thm for hat C}.
\qed

\section{Conclusion}

We decompose the group algebra of a finite dihedral group 
of order $2n$ with $n$ being odd over any finite filed $F$ 
into an orthogonal direct sum of a special component of dimension $2$ and 
some $2\times 2$ matrix algebras. 
With the structure we find two kinds of random dihedral group codes. 
The random dihedral codes are self-dual, or maximal self-orthogonal, 
or LCD under different conditions.  
And the random dihedral codes have nice asymptotic behavior so that,  
suitably choosing a positive real number~$\delta$ 
and the code lengths $2n_1, 2n_2, \cdots$ going to infinity, 
we proved that the probability for the relative minimum distance 
of the random dihedral codes greater than $\delta$ is convergent to $1$. 

As consequences, if ${\rm char}\,F=2$, 
then self-dual dihedral codes are asymptotically good.
In the case that ${\rm char}\,F$ is odd,
there exist asymptotically good maximal self-orthogonal dihedral codes of rate
tending to $\frac{1}{2}$; and, 
LCD dihedral codes of rate $\frac{1}{2}$ are asymptotically good.

In the case that ${\rm char}\,F=2$, 
both the two kinds of random dihedral codes we discussed are not LCD.
Though the LCD dihedral codes  
exist in that case, for the moment 
we have no good idea to study their asymptotic behavior, 
and we guess that they probably have no good asymptotic property.   

\section*{Acknowledgements}
Thanks are given to Alahmadi, \"Ozdemir, Sol\'e 
and Borello, Willems for showing us 
their interesting works \cite{AOS} and \cite{BW}.
We are grateful to the editor and the anonymous referees 
for taking time to read and comment on the paper very carefully.
Their nice comments and suggestions 
helped us to improve the paper very much.



\begin{thebibliography}{99}

\bibitem{AOS} A. Alahmadi, F. \"Ozdemir, P. Sol\'e,
``On self-dual double circulant codes'',
{\it Des. Codes Cryptogr.}, vol. 86, pp. 1257-1265, 2018.

\bibitem{AKS} S. A. Aly, A. Klappenecker, P. K. Sarvepalli, 
``Duadic group algebra codes'', 
 ISIT2007, Nice, France, June 24-June 29, pp. 2096-2100, 2007.

\bibitem{BF} A. Barg and G. D. Forney, 
``Random codes: Minimum distances and error exponents'', 
{\it IEEE Trans. Inform. Theory}, vol. 48, no. 9, pp. 2568-2573, 2002.

\bibitem{BM}L. M. J. Bazzi, S. K. Mitter,
``Some randomized code constructions from group actions'',
{\it IEEE Trans. Inform. Theory}, vol. 52, pp. 3210-3219, 2006.

\bibitem{BW} M. Borello, W. Willems, 
``Group codes over fields are asymptotically good'',
Finite Fields and Their Applications, 
vol. 68(Dec), 2020, 101738.

\bibitem{CPW}
C. L. Chen, W. W. Peterson, E. J. Weldon, ``Some results on quasi-cyclic codes'',
{\it Information and Control}, vol. 15, pp. 407-423, 1969.

\bibitem{C}V. Chepyzhov, ``New lower bounds for minimum distance of linear
quasi-cyclic and almost linear quasi-cyclic codes'',
{\it Problem Peredachi Informatsii}, vol. 28, pp. 33-44, 1992.

\bibitem{D} B. K. Dey, ``On existence of good self-dual quasi-cyclic codes'',
{\it IEEE Trans. Inform. Theory}, vol. 50, pp.1794-1798, 2004.

\bibitem{FL} Yun Fan, Liren Lin, 
``Thresholds of random quasi-abelian codes'',
{\it IEEE Trans. Inform. Theory}, vol. 61, no. 1, pp. 82-90, 2015.

\bibitem{FY} Yun Fan, Yuan Yuan, 
``On Self-dual Permutation Codes'',
{\it Acta Mathematica Scientia}, vol. 28B, no. 3, pp. 633-638, 2008.

\bibitem{G}
N. E. Gilbert, ``A comparison of signalling alphabets'', 
{\it Bell Sys. Tech. Journal}, vol. 31, pp. 504-522, 1952.

\bibitem{Hasse}Helmut Hasse, 
``\"Uber die Dichte der primzahlen $p$, f\"ur die eine vorgegebene 
ganzrationale zahl $a\ne 0$ von gerader bzw. ungerader ordnung mod p ist'', 
{\it Math. Annalen}, vol. 166, pp. 19-23, 1966.

\bibitem{HP} W. C. Huffman, V. Pless, 
{\it Fundamentals of Error-Correcting Codes}, 
Cambridge University Press, 2003.

\bibitem{HB} B. Huppert, 
{\it Endliche Grouppen I}, Springer Verlag, Berlin Heidelberg New York, 1967.

\bibitem{IW} K. A. Schouhamer Immink and J. H. Weber, 
``Very efficient balanced codes'', 
{\it IEEE J. Sel. Areas Commun.}, vol. 28, no. 2, pp.188-192, 2010.

\bibitem{K}T. Kasami,
``A Gilbert-Varshamov bound for quasi-cyclic codes of rate 1/2'',
{\it IEEE Trans. Inform. Theory}, vol. 20, pp. 679, 1974.

\bibitem{KR} L. Kathuria, M. Raka, 
``Existence of cyclic self-orthogonal codes: 
 A note on a result of Vera Pless'', 
{\it Adv. Math. Commun.}, vol. 6, pp. 499-503, 2012.

\bibitem{LS} San Ling, P. Sol\'e,
``Good self-dual quasi-cyclic codes exist'',
{\it IEEE Trans. Inform. Theory}, vol. 49, pp. 1052-1053, 2003.

\bibitem{MW06} C. Mart\'inez-P\'erez, W. Willems,
``Is the class of cyclic codes asymptotically good?''
{\it IEEE Trans. Inform. Theory}, vol. 52, no. 2, pp. 696-700, 2006.

\bibitem{MW} C. Mart\'inez-P\'erez, W. Willems,
``Self-dual double-even $2$-quasi-cyclic transitive codes
are asymptotically good'',
{\it IEEE Trans. Inform. Theory}, vol. 53, pp. 4302-4308, 2007.

\bibitem{M74} J. L. Massey,
``On the fractional weight of distinct binary n-tuples'',
{\it IEEE Trans. Inform. Theory}, vol. 20, pp. 130, 1974.



\bibitem{MU} M. Mitzenmacher, E. Upfal,
{\it Probability and Computing:
  Randomized Algorithm and Probabilistic Analysis},
   Cambridge Univ. Press, Cambridge, 2005.

\bibitem{O} R. W. K. Odoni, ``A conjecture of Krishnamurthy
on decimal periods and some allied problems'', 
{\it J. of Number Theory}, vol. 13, pp. 303-319, 1981.

\bibitem{P67} J. N. Pierce,
``Limit distribution of the minimum distance of random linear codes'', 
{\it IEEE Trans. Inform. Theory}, vol.13, pp. 595-599, 1967.

\bibitem{P85} P. H. Piret,
``An upper bound on the weight distribution of some codes'',
{\it IEEE Trans. Inform. Theory}, vol. 31, pp. 520-521, 1985.


\bibitem{S86} I. E. Shparlinsky,
``On weight enumerators of some codes'',
{\it Problemy Peredechi Inform.}, vol. 22, no.2, pp. 43-48, 1986.

\bibitem{V} R. R. Varshamov,
``Estimate of the number of signals in error-correcting codes''
(in Russian), {\it Dokl. Acad. Nauk}, vol. 117, no.5, pp. 739-741, 1957.

\bibitem{W} W. Willems,
``A note on self-dual group codes'',
{\it IEEE Trans. Inform. Theory}, vol. 48, pp. 3107-3109, 2002.

\end{thebibliography}
\end{document}